\documentclass[10pt, conference, compsocconf]{IEEEtran}
\usepackage{multirow}
\usepackage{graphics}
\usepackage{pifont}
\usepackage{subfig}
\usepackage{footmisc}
\usepackage{threeparttable}

\usepackage{cite}

\ifCLASSINFOpdf
 \usepackage[pdftex]{graphicx}
\else
\fi
\begin{document}

\title{C2MS: Dynamic Monitoring and Management of Cloud Infrastructures}



%
\author{\IEEEauthorblockN{Gary A. McGilvary\IEEEauthorrefmark{1},
Josep Rius\IEEEauthorrefmark{2},
\'I\~{n}igo Goiri\IEEEauthorrefmark{3} and 
Francesc Solsona\IEEEauthorrefmark{4} and
Adam Barker \IEEEauthorrefmark{5} and 
Malcolm Atkinson \IEEEauthorrefmark{1}}
\IEEEauthorblockA{\IEEEauthorrefmark{1}Edinburgh Data-Intensive Research Group, 
School of Informatics, The University of Edinburgh 
\\ Email: gary.mcgilvary@ed.ac.uk}
\IEEEauthorblockA{\IEEEauthorrefmark{2} ICG Software, Industrial Polygon, Mestral St., 25123 Torrefarrera, Spain.}
\IEEEauthorblockA{\IEEEauthorrefmark{3} Dept. of Computer Science, Rutgers University.}
\IEEEauthorblockA{\IEEEauthorrefmark{4} Dept. of Computer Science and Industrial Engineering, University of Lleida.}
\IEEEauthorblockA{\IEEEauthorrefmark{5}School of Computer Science, University of St Andrews.}}

\maketitle
\begin{abstract}
Server clustering is a common design principle employed by many organisations who require high availability, scalability and easier management of their infrastructure. Servers are typically clustered according to the service they provide whether it be the application(s) installed, the role of the server or server accessibility for example. In order to optimize performance, manage load and maintain availability, servers may migrate from one cluster group to another making it difficult for server monitoring tools to continuously monitor these dynamically changing groups. Server monitoring tools are usually statically configured and with any change of group membership requires manual reconfiguration; an unreasonable task to undertake on large-scale cloud infrastructures.

In this paper we present the Cloudlet Control and Management System (C2MS); a system for monitoring and controlling dynamic groups of physical or virtual servers within cloud infrastructures. The C2MS extends Ganglia - an open source scalable system performance monitoring tool - by allowing system administrators to define, monitor and modify server groups without the need for server reconfiguration. In turn administrators can easily monitor group and individual server metrics on large-scale dynamic cloud infrastructures where roles of servers may change frequently. Furthermore, we complement group monitoring with a control element allowing administrator-specified actions to be performed over servers within service groups as well as introduce further customized monitoring metrics. This paper outlines the design, implementation and evaluation of the C2MS. 

\end{abstract}


\IEEEpeerreviewmaketitle

\section{Introduction}
System monitoring is a critical component in the operation of any infrastructure, involving the observation and analysis of infrastructure state, to detecting and recovering from faults or unexpected downtime. Monitoring is especially useful in complex systems when administrators are not able to understand their infrastructure easily. 

Organizations who wish to reduce the complexity of their infrastructure as well as maintain high availability and scalability, can employ server clustering; the grouping of servers based on administrator-defined characteristics. Servers may be clustered according to the service they provide such as the applications they execute, the software installed, the server's role or server accessibility to clients, for example.

Regardless of the type of clustering employed, monitoring these groups of servers within cloud infrastructures can be troublesome if servers are migrated from one group to another, for example to optimise infrastructure performance and maintain service availability. Typically monitoring tools are statically configured hence any membership change requires an administrator to:

\begin{itemize}
\item manually change the monitoring tool's configuration file to specify the new group the server belongs to. A system configuration tool could be used, however this increases the complexity and effort needed to monitor the infrastructure. 
\item restart the monitoring process upon the server being monitored. This is required for the server to adopt the changes.
\item restart the data collection process (located on a central server); a process that may take several minutes for the changes to take effect. This therefore restricts the use of the monitoring tool during this period.
\end{itemize}

\noindent Within a large dynamic cloud infrastructure where servers frequently migrate between groups, the manual effort to perform these tasks outweigh the desired result and prevent the administrator from concentrating on more important tasks. Many monitoring tools exist for infrastructures whose servers need little or no configuration changes over their lifetime however none are designed to monitor dynamically changing groups of servers; others have noted the lack of tools for rapidly changing environments, particularly within cloud environments \cite{M.Lindner2010} \cite{Ward2012}. The lack of dynamic tools has contributed to developers limiting tool complexity however because server clustering is commonly employed by organizations, a large subset will not be able to easily and effectively monitor their dynamically clustered cloud infrastructure. 

To solve the aforementioned problems, this paper proposes the Cloudlet Control and Monitoring System (C2MS); the dynamic groups of servers are what we define as "cloudlets".  

The C2MS makes the following major contributions: 

\begin{itemize}
\item allows administrators to create cloudlets and migrate servers between cloudlets via the C2MS interface.
\item monitors dynamically changing groups of servers by extending Ganglia \cite{ganglia}\cite{wide_area_ganglia} \cite{Massie2012} to avoid the re-configuration of servers and restart of monitoring process when they migrate between cloudlets. 
\end{itemize}

\noindent Administrators are then able to define and monitor the overall state of cloudlets independently without explicit configuration changes, in turn making it easy to view monitoring data tailored towards the architecture of the system. This is an innovation that overcomes the time-consuming limitations of previous monitoring tools in turn freeing administrators of large-scale systems to focus on operational challenges with improved information. In addition to our major contribution, we have made a series of extensions/improvements to Ganglia with the aim of making infrastructure monitoring and management easier for administrators. We introduce:

\begin{itemize}
\item further metrics than those provided by Ganglia that are commonly monitored nowadays; these are power usage and CPU temperature monitoring.
\item a management element on top of Ganglia to give administrators the ability to quickly take control of individual servers or entire cloudlets by issuing administrator-specified commands. This may be used for upgrading existing software or installing new software over many servers for example.
\end{itemize}

\noindent A large number of server monitoring and management tools exist independently however very few provide both these functions. As such we also reduce the effort required for installation and maintenance of these independent packages by combining these features into a single tool. The C2MS can be used on a number of infrastructures such as clusters, clouds, grids and is available to download online at \cite{c2ms_gary}.

In this paper, we discuss the how the C2MS offers the aforementioned features in detail and how they are implemented. The rest of this paper is organized as follows: next we give an overview of related research and then provide a system overview of the C2MS in Section 3. Section 4 describes our implementation where we then evaluate our C2MS in Section 5. Finally we conclude with a summary of our tool and describe future work in Section 6.

\section{Related Work}
The number of system monitoring and management tools are plentiful however none are able to monitor dynamically changing groups of servers without the need for reconfiguration of servers. We outline some of the current leading monitoring and management tools in the field while paying specific attention to the different features offered by our work. 

The C2MS uses Ganglia as its foundation for infrastructure monitoring. Ganglia is a popular open source scalable system performance monitoring tool \cite{ganglia,wide_area_ganglia,monitoring_taxonomy} used widely in the High Performance Computing (HPC) community. Its popularity, easy installation process, easy to use web interface and its extensibility were the main factors why we chose Ganglia to build the C2MS upon.

The Ganglia framework relies on two daemons: \textit{gmond} and \textit{gmeta} \cite{Massie2012}. For brevity, the \textit{gmond} daemon collects resource usage information about the host it runs upon (remote server) and sends periodic heartbeat messages via a UDP multicast protocol to the entire Ganglia cluster. The \textit{gmeta} daemon collects the aggregated XML and exports it to the PHP web interface hosted on a central server. 

Ganglia monitors different groups of machines (or Ganglia `clusters') by allowing the administrator to define the cluster/group name within the remote server's configuration file. This allows the PHP web interface to display the aggregated data for this group. Any remote server changing to an alternative group requires manual reconfiguration and the restart of the \textit{gmond} and \textit{gmeta} daemons on the remote server and central server respectively. The C2MS offers an abstract layer built on top of Ganglia allowing group membership changes without the need for reconfiguration or daemon restarts upon any server.

Another popular monitoring tool is Nagios describing itself as \textit{the industry standard in IT infrastructure monitoring} \cite{nagios}. Unlike Ganglia, it has two powerful features: administrator alerting and response to failures. Administrators are notified of any infrastructure problems and can define event handlers to respond to problems automatically. However, similar to Ganglia, Nagios monitors system metrics, network protocols, servers, and network infrastructure, is scalable to thousands of servers, stores historical data and is open source. 

Nagios does monitor groups of servers however with the main purpose of configuration simplification and making navigation via the Nagios GUI easier; this is not for monitoring dynamic groups of servers. Nagios is also configured by modifying a number of configuration files however located on a central server rather than on remote servers. The configuration files define all the remote servers and the operations to be performed for example, check availability, monitor server resource usage etc. As a result, any modifications to the configuration requires Nagios to restart and as such, the statically configured Nagios makes monitoring a dynamic group of servers within a cloud difficult; the C2MS provides this functionality.

ScaleExtreme is an all-in-one server monitoring and management tool primarily designed to run on cloud platforms \cite{scalextreme}. ScaleExtreme monitors CPUs, disks, memory, network, processes and application specific metrics for Apache, MySQL, MongoDB and many others. The server management features include a browser-based command shell however commands cannot be executed over multiple machines concurrently. Furthermore ScaleExtreme allows servers to be grouped however only for the purposes of easy management. A feature rich version of ScaleExtreme exists but is charged at a monthly rate. With fewer features, the free trial version can be used but the collected data can only be used for one year. ScaleExtreme also does not offer the dynamic monitoring behaviour we aim to solve. 

Wright \textit{et al} outline their view of a dynamic cloud management and monitoring system tailored towards services, e.g. a web server \cite{Wright2010}. The authors also note the lack of dynamic tools for cloud environments and in turn have created the Cloud Management System (CMS). The CMS is similar to the C2MS and Ganglia in many ways, however their implementation monitors services running on cloud instances rather than the instances themselves. The C2MS provides this functionality in a dynamic context when servers are sporadically available and migrate between cloudlets dependent on an organization's infrastructure requirements.

Birman \textit{et al} outline their distributed self-configuring monitoring and adaptation tool called Astrolabe \cite{Birman2003}. Astrolabe works like any other monitoring tool by observing the state of an infrastructure where the tool is installed.  However it differs by essentially creating a virtual system-wide hierarchical relational database based on a peer-to-peer protocol meaning no central server needs to exist to collect monitoring data. By performing distributed data analysis, Astrolabe can create performance summaries of \textit{zones} --- machines typically grouped based on the shortest latency between one another or simply an administrator-specified group --- by data aggregation; a method we use to create graphs of cloudlets.

A major advantage of Astrolabe is its ability to adapt to configuration changes without the need of restarting the monitoring tool however administrator-specified groups and the resources to be monitored need to be manually configured in their \textit{configuration certificate} file; once again a troublesome task for a large dynamically changing cloud infrastructure. 

We have now outlined similar research in the area and shown how the functionality and purpose of the C2MS is distinct. We believe the C2MS fills a gap in the market where statically configured monitoring tools cannot monitor a dynamically changing cloud infrastructure. Now we give an overview and implementation details of the C2MS.

\section{System Overview}
The C2MS consists of three major components: the \textit{Monitoring}, \textit{Control} and \textit{Cloudlet Creator} components shown in Figure 1. These components relate to the web pages that an administrator can navigate to. 
\vspace{-1em}
\begin{figure}[h!]
\centering
\includegraphics[width=0.45\textwidth]{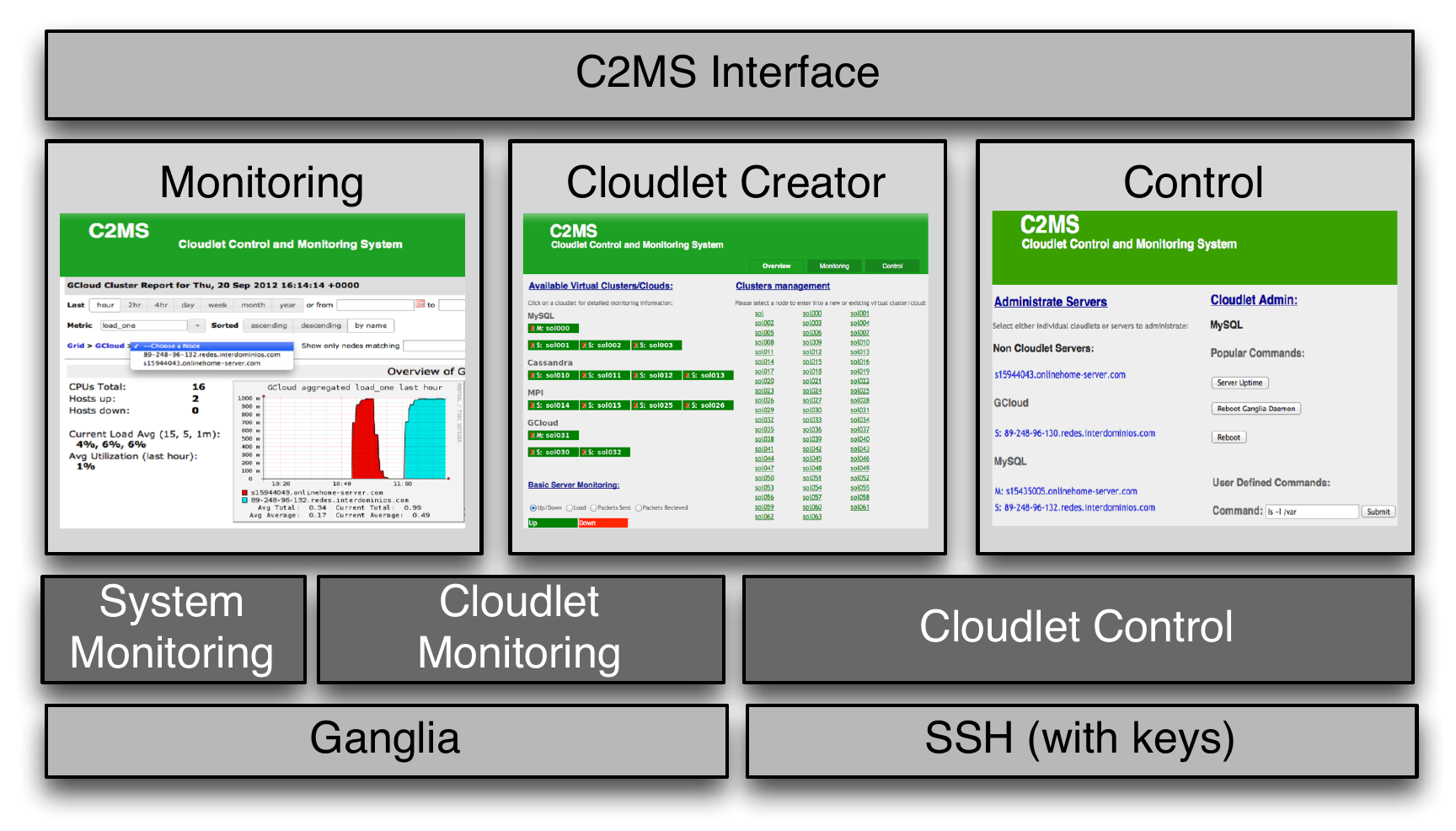}
\caption{The C2MS Architecture} 
\end{figure}

The Monitoring component is a modified version of Ganglia allowing individual, cloudlet or entire system monitoring where the former and latter are provided by Ganglia by default. The Control component gives administrators the ability to control either single machines or entire cloudlets via SSH. Both components use the output from the Cloudlet Creator to determine the servers within a cloudlet. The component functionalities are easily accessed through the C2MS interface via the tabs Overview, Monitoring and Control as shown in Figure 2.
\vspace{-0.4em}
\begin{figure}[ht]
\centering
\includegraphics[width=0.51\textwidth]{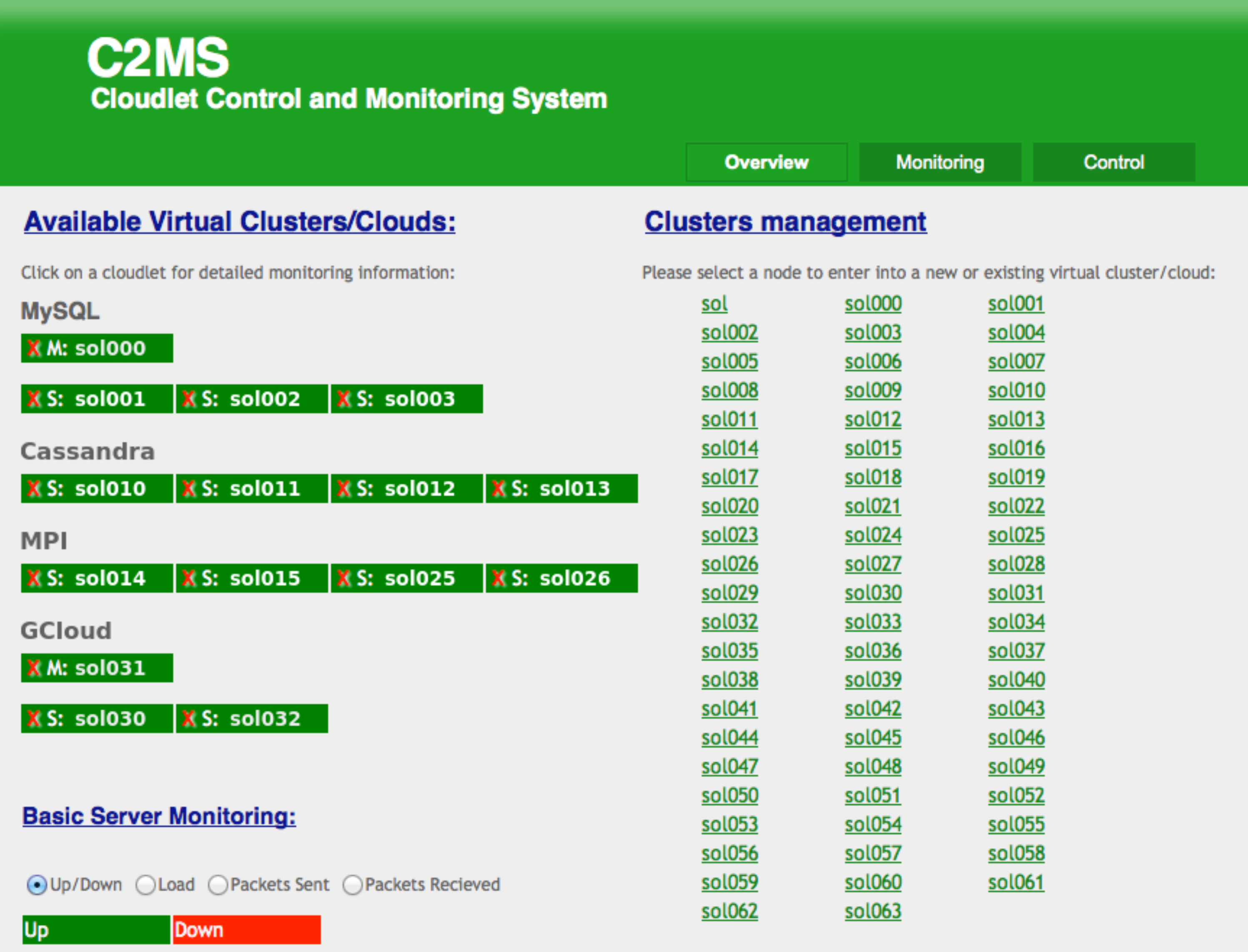}
\caption{C2MS PHP Interface Overview Page}
\end{figure}

The Overview tab displays all servers within the system that have the \textit{gmond} daemon running and allows cloudlets to be created for combined monitoring and control. Administrators can create cloudlets by selecting a server from the list of all those available (right) and entering the desired cloudlet name on a pop-up text field. If the cloudlet exists, the server is successfully added, else a new cloudlet is created with the selected server; the created cloudlets are shown on the left hand side of the Figure 2, e.g. MySQL, MPI. The administrator is then able to view cloudlet or system specific graphs by clicking on the cloudlet name or via the Monitoring tab. 

To remove a server from a cloudlet, an administrator is required to click on the `X' marked besides the server name. Also administrators are able to view basic monitoring characteristics (bottom left) such as whether each server is up/down and the CPU and network load based on the check button selected; this information is displayed via color-coded servers. 

The Control tab provides a similar page allowing administrator-defined commands to be executed either on individual machines or over entire cloudlets. Typically, Ganglia allows public users to view server resource usage data however because this tool is intended for private use (i.e., administrators only) --- due to the Cloudlet Creator and control elements --- we provide a login page with changeable credentials to prevent public users accessing the system and performing malicious tasks. We leave it to the administrator to provide additional security measures if required.
 
\section{Implementation}
We now explore how the functionality behind the interface components are implemented and integrated to provide a dynamic monitoring and control system.

\begin{figure*}[ht]
  \begin{center}
\includegraphics[width=0.99\textwidth]{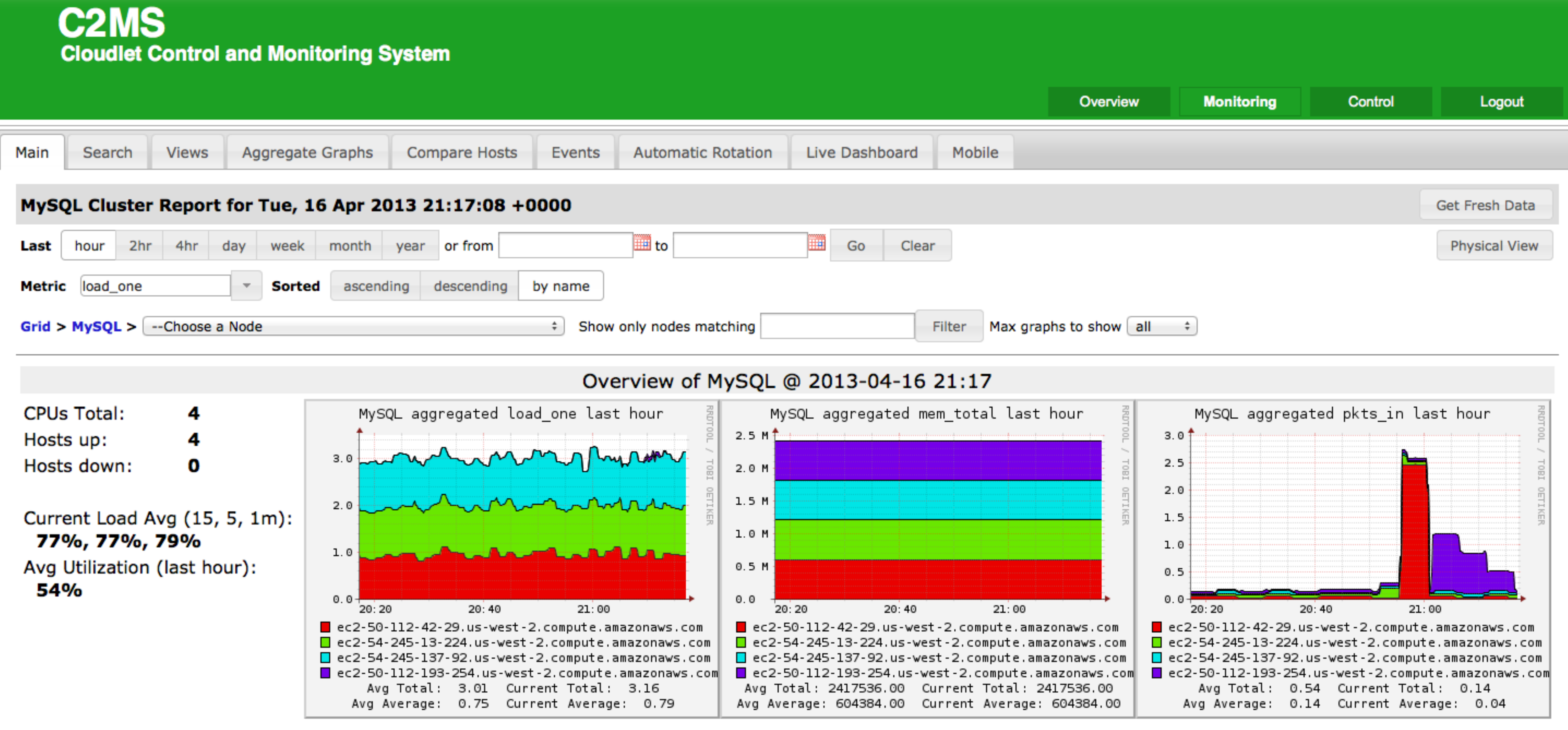}
  \end{center}
 \caption{Monitoring Data of a Cloudlet}
\end{figure*}

\subsection{Creating Cloudlets}
In order to monitor and view the state of an entire cloudlet, the C2MS must be made aware that a cloudlet exists and the servers it contains. Upon installing and configuring Ganglia, an administrator simply needs to modify each server's \textit{gmond} configuration file to include the hostname or IP address of the Ganglia interface and specify the Ganglia cluster as \textit{Initial}. 

When the Ganglia interface receives monitoring data from these servers, the C2MS will register that the servers are present and display them to the user as shown in the right hand side of Figure 2. By giving each remote server the same Ganglia cluster name, we can virtually partition this single group of servers at a higher level to allow cloudlets to be created. Details on how to configure a Ganglia cluster can obtained from \cite{ganglia_website} and instructions on how to setup the C2MS are present within the C2MS downloadable file \cite{c2ms_gary}.

When an administrator creates a cloudlet via the C2MS interface, the server and cloudlet name specified is recorded in a file named \textit{clusters} within the \textit{/etc/ganglia/} folder; this file contains a list of cloudlets and their member servers. The C2MS interface then  displays this grouping by reading and parsing the \textit{clusters} file. However at this point, Ganglia will not display cloudlet based monitoring data as it is unaware that a cloudlet or a number of them exist. To enable cloudlet based monitoring, Ganglia requires that each Ganglia cluster has a directory present in \textit{/var/lib/ganglia/rrds/}, named after the cloudlet, containing directories for individual remote servers; these directories contain monitoring data (\textit{.rrd} files) for the server. 

The stored \textit{.rrd} files are created by RRDtool (Round Robin Database); an open source tool for data logging and graphing historical data between specified times. Upon cloudlet creation, the C2MS creates the appropriate cloudlet folder within the \textit{/var/lib/ganglia/rrds/} directory and links to the original \textit{.rrd} files of each server within the \textit{Initial} folder. We therefore do not need to replicate any data which would in turn introduce overheads. Hence with the creation of a new cloudlet, Ganglia is lead to believe that it has received monitoring data from a new cluster which contains the servers listed in the \textit{/var/lib/ganglia/rrds/cloudlet\_name} directory.

In the event of cloudlet creation, deletion or a change of a server's cloudlet membership from one to another, the C2MS only needs to modify configuration files linked to our tool and none that are related to the operation of Ganglia; this allows us to avoid restarting the Ganglia daemons. These configuration changes are obscured from the administrator and are automatically performed by the C2MS interface. 

\subsection{Monitoring Cloudlets}
The information we are interested in displaying to the administrator is the entire state of multiple cloudlets via summary graphs per Ganglia metric. Each page displaying monitoring data of a cloudlet allows users to either view a summary of the current cloudlet state or select individual servers to examine their resource usage in more detail. Figure 3 shows both these features which are inherited from Ganglia.

Firstly, we see that four servers exist within the `MySQL' cloudlet, both from the number of `hosts up' and the total of CPUs. The graphs shown are only specific to the `MySQL' cloudlet with colours making the distinction between individual servers present in the cloudlet. To create cloudlet summary graphs, data aggregation is used and this is apparent in the graphs above where data from one cloud server is stacked upon another, in turn displaying the total resource use for the selected cloudlet; different cloudlets can be selected on the `Overview' page of Figure 2. 

The depicted graphs automatically change when servers are added to or removed from the cloudlet. To create aggregated graphs dynamically, Ganglia calls the file \textit{/var/www/ganglia-web/stacked.php} when the page is viewed; a default file of the Ganglia implementation. This has been modified to only create stacked graphs for servers present in a cloudlet rather than an entire system as regular Ganglia would do; the same has been applied to the number of `hosts up', `hosts down', and `CPUs Total'. The PHP file returns PNG files of the created graphs and these are displayed via the Ganglia interface.

Graph data aggregation can be easily achieved through the use of RRDtool. We implement this through PHP calls to RRDtool however this can be easily explained by the use of \textit{rrdtool}'s graph function shown below.\\

\noindent\textit{rrdtool graph agg\_graph.png \textendash \textendash imgformat=PNG}\\
\textit{DEF:one=server1\_metric.rrd:sum:AVERAGE}\\
\textit{AREA:one\#00CF06::STACK}

\noindent\textit{DEF:two=server2\_metric.rrd:sum:AVERAGE}\\
\textit{AREA:two\#CC0000::STACK \textendash \textendash start timeX \textendash \textendash end timeY}\\

First we define variables, one for each of the server's \textit{.rrd} files to be aggregated (e.g. \textit{one} and \textit{two}). The data \textit{sum} is then plotted using average values. We use the AREA shape to plot the variable values with different colours and in the form of a STACK where one dataset is placed on top of another. We also enter a start and end time specified by the administrator to allow historical cloudlet monitoring data to be accessed. Other arguments are omitted here for clarity that relate to the appearance of graphs such as the width, height and labels.

\subsection{Additional Metrics}
The C2MS not only measures basic resource usage such as CPU, memory, etc, but by installing additional modules, one can also monitor power consumption and temperature.

Monitoring temperature requires a server's CPU(s) to possess built-in temperature monitoring capabilities such as those found in Intel Core based processors and others \cite{intel_temp}. The C2MS collects temperature data by adding a monitoring module to the \textit{gmond} daemon of every cloud server which periodically polls the CPU's Digital Thermal Sensor to obtain temperature data. This data is then available to the \textit{gmeta} daemon which in turn can display this information. Similarly, this information can be aggregated to show data for single cloudlets or for single servers. Figure 4 shows one other method we use for displaying this data where servers are presented as a heat map in the rack format.

\begin{figure}[h!]
  \begin{center}
\includegraphics[width=0.35\textwidth]{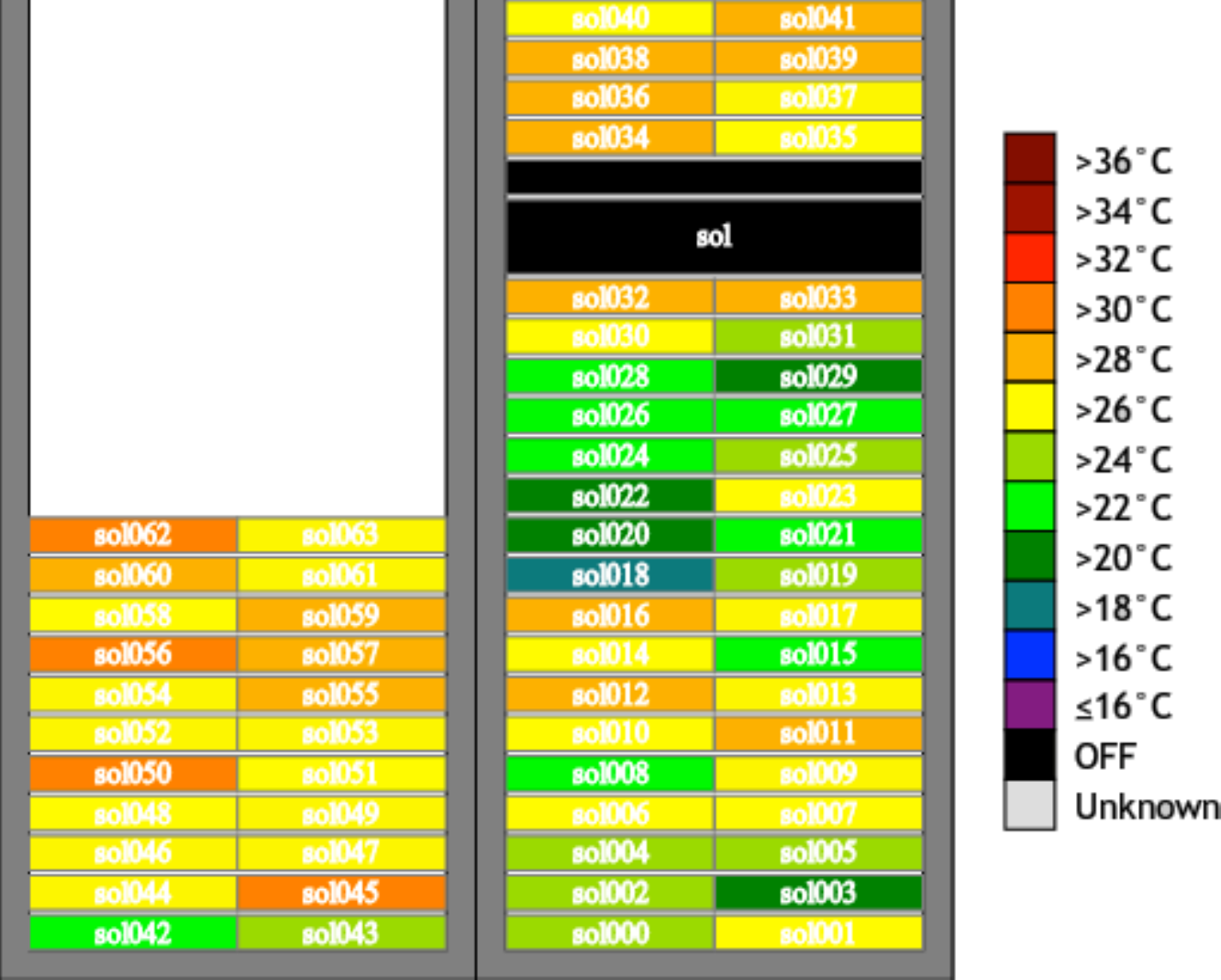}
  \end{center}
 \caption{CPU Temperature Data Output}
\end{figure}

Similar to temperature monitoring, power observation requires the appropriate power monitoring hardware or a Power Distribution Unit (PDU). The data recorded by the PDU is then periodically queried and stored in RRD files following the Ganglia RRD structure. These are then exported to graphs and added to the Ganglia interface for viewing. As the purpose of our tool is to dynamically monitor sets of cloud servers, we also allow power consumption to be monitored for cloudlets by graph aggregation. To distinguish power usage for servers connected to the same PDU, the administrator must identify each PDU and it's connected servers in a file accessed by the C2MS. These details include the server name, the MAC address as well as the PDU identifier and which outlet the server is connected to. The tool therefore allows per server or per cloudlet power usage monitoring.

\begin{figure*}[ht]
  \begin{center}
\includegraphics[width=0.99\textwidth]{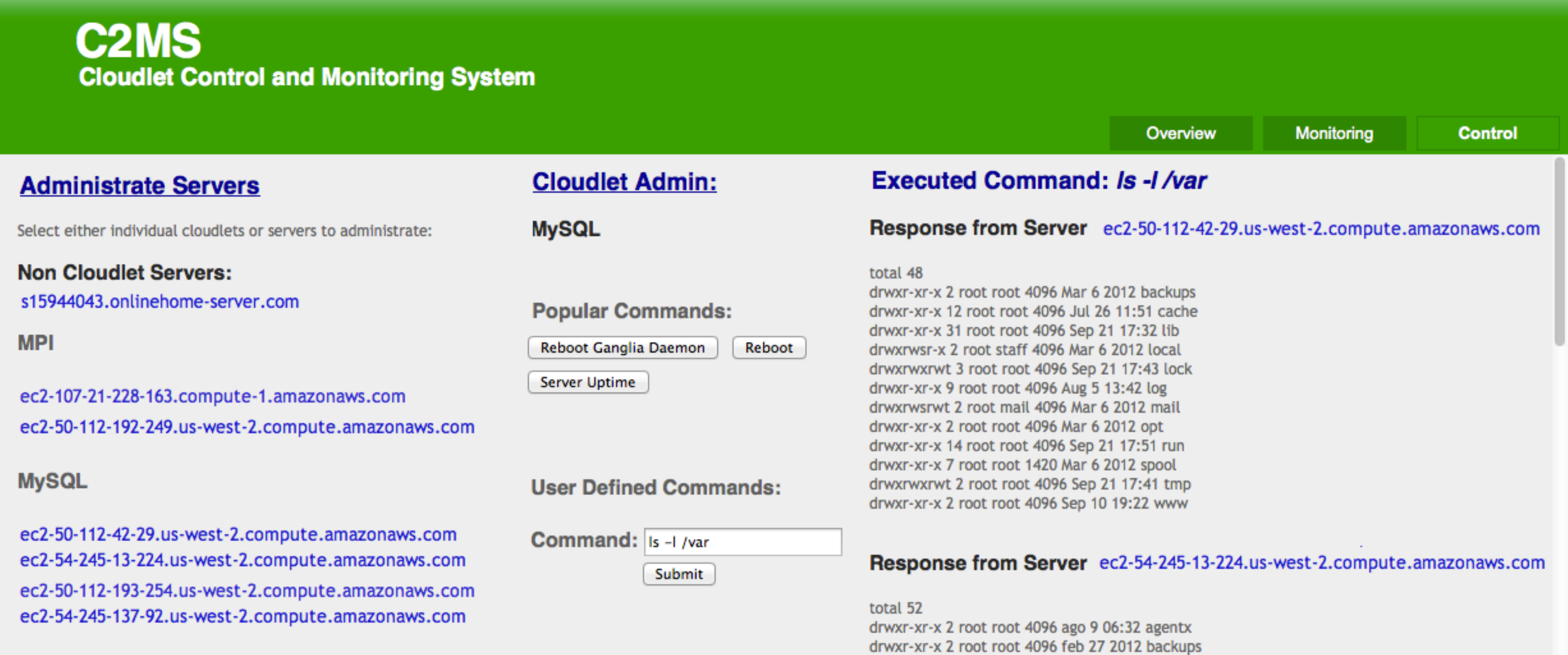}
  \end{center}
 \caption{Controlling a Cloudlet}
\end{figure*}

\subsection{Controlling Cloudlets}
Our final contribution incorporates a server management component into the C2MS. Administrators are not only able to control individual servers but can issue specified instructions over cloudlets or individual servers. The instructions may be input manually or selected from a list of popular commands, as shown in Figure 5, where the MySQL cloudlet is selected. In order to introduce this functionality, we investigated a number of popular tools to determine whether they satisfied our requirements for use within the C2MS. Such a tool must:

\begin{enumerate}
\item allow the grouping of servers and concurrent command execution upon these groups.
\item not require the installation of software on remote servers within cloudlets.
\item be easy to integrate into the C2MS.
\end{enumerate}

\noindent The tools we investigated were: Webmin, Capistrano and \textit{cexec}.

Firstly, Webmin is a browser-based system administration tool for Unix \cite{webmin}. Webmin allows the grouping of servers into cloudlets and commands to be executed per-cloudlet. However Webmin requires the installation of software on remote servers and the integration process of Webmin into the C2MS would not be simple as it would have to be modified. For example, the creation of a cloudlet via the C2MS interface would have to be reflected in the Webmin service to avoid administrators creating a cloudlet twice via the monitoring and control components.

Secondly, Capistrano is an open source tool for running scripts and commands concurrently over multiple servers. It allows the grouping of servers by simply specifying these groups within their configuration \textit{capfile}; this therefore can be easily accessed and modified by the C2MS. Furthermore, Capistrano does not require any installation of software on remote servers meaning it satisfies all requirements. Capistrano does however use SSH and assumes that SSH keys are exchanged between the central and remote servers to allow password-less login for commands to execute \cite{ssh_passwordless}. This is in contrast to Webmin which uses the software installed on remote servers to create tunnels to send instructions over.

Finally, \textit{cexec} is cluster tool that simply executes commands over multiple servers concurrently \cite{PeterKacsuk2004}. To execute a command over a set of servers, \textit{cexec} requires that a configuration file exists listing the hostname or IP address of the servers in a cloudlet alongside the cloudlet name; multiple cloudlets can exist allowing the administrator to specify the cloudlet to execute the command over. Like Capistrano, we can automatically generate this file by entering the hostnames of the cloudlet members, taken from the \textit{/etc/ganglia/clusters} file, into \textit{cexec}'s configuration file, making integration into the C2MS easy. Furthermore, \textit{cexec} does not require any software installation on target machines as it uses SSH and assumes SSH keys are exchanged between servers.

The C2MS currently uses \textit{cexec} as its control component. This is based on the simplicity of the tool as well as its performance as explored in the following Section.

\begin{figure*}
\centering
\subfloat[]{\includegraphics[width=3.15in]{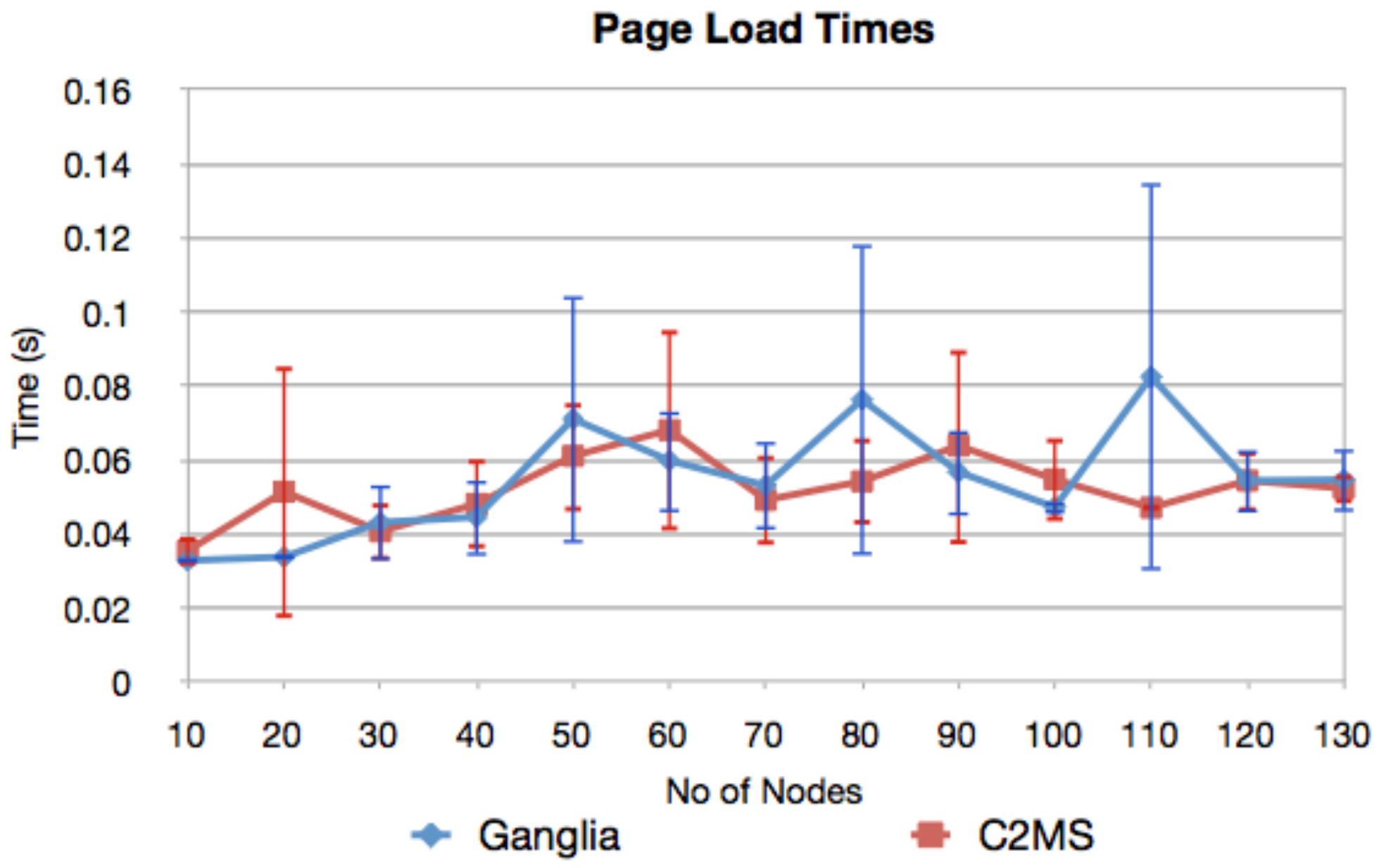}}
\subfloat[]{\includegraphics[width=2.9in]{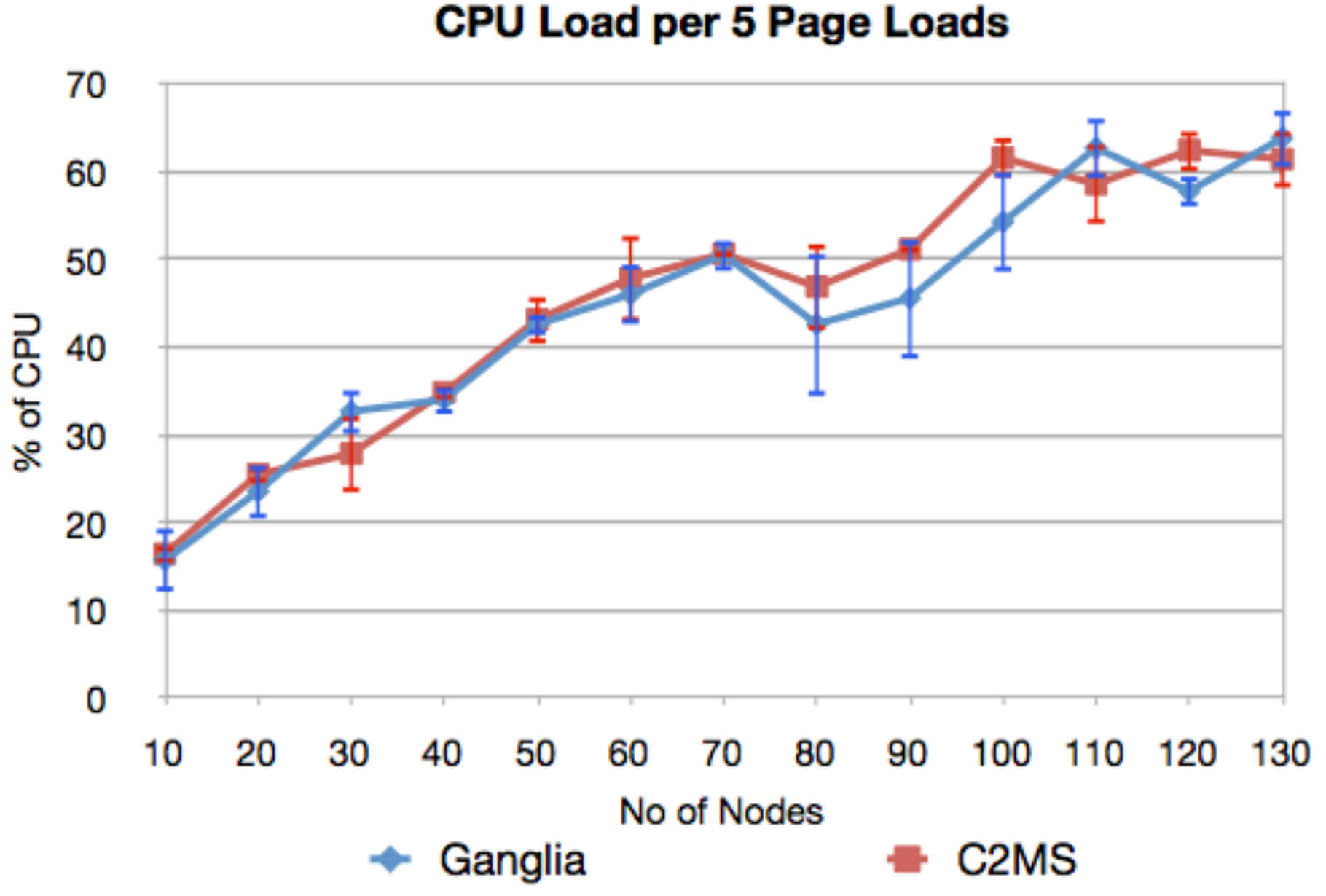}}
\caption{Load Comparison of Ganglia and the C2MS}
\end{figure*}

\section{Evaluation}
Ganglia is commonly used in the HPC and Grid communities where clusters, like cloud infrastructures, typically contain a large number of servers. We now investigate how effectively the C2MS can monitor such systems by determining whether our implementation introduces any additional overhead above that already introduced by regular Ganglia. We then determine the optimal method of server management and if the C2MS can execute administrator commands over a large number of machines quickly. We perform these experiments on Amazon EC2 with the Ganglia/C2MS interface running on a Large Ubuntu 12.04 instance and remote servers running on Micro instances of the same type.

\subsection{Monitoring Performance}
Ganglia is well known for its scalable implementation hence the modifications we have made must also be able to cope with an increase in the number of servers. We use at most 130 servers; the maximum number of instances we could instantiate on Amazon EC2. First we test if our method of graph aggregation and the operations that underpin it introduce any overhead when compared to regular Ganglia. To test this, we split the experiment into two parts: one to record the page load times of both systems and another to determine the impact on the Apache server displaying the data. 

\subsubsection{Page Load Times}
We first explore whether viewing a cloudlet's monitoring output via the C2MS takes additional time to load when compared to regular Ganglia. For example, if we view the monitoring output of a 50 node Ganglia cluster, does the C2MS introduce any overhead when we view a cloudlet of the same size? We compare the page load times of an increasing Ganglia cluster and C2MS cloudlet size. By adding a simple PHP page load counter to the page displaying monitoring data, loading each page 15 times and taking the average value, we obtain the results shown in Figure 6(a).

\begin{figure*}
\centering
\subfloat[]{\includegraphics[width=3.1in]{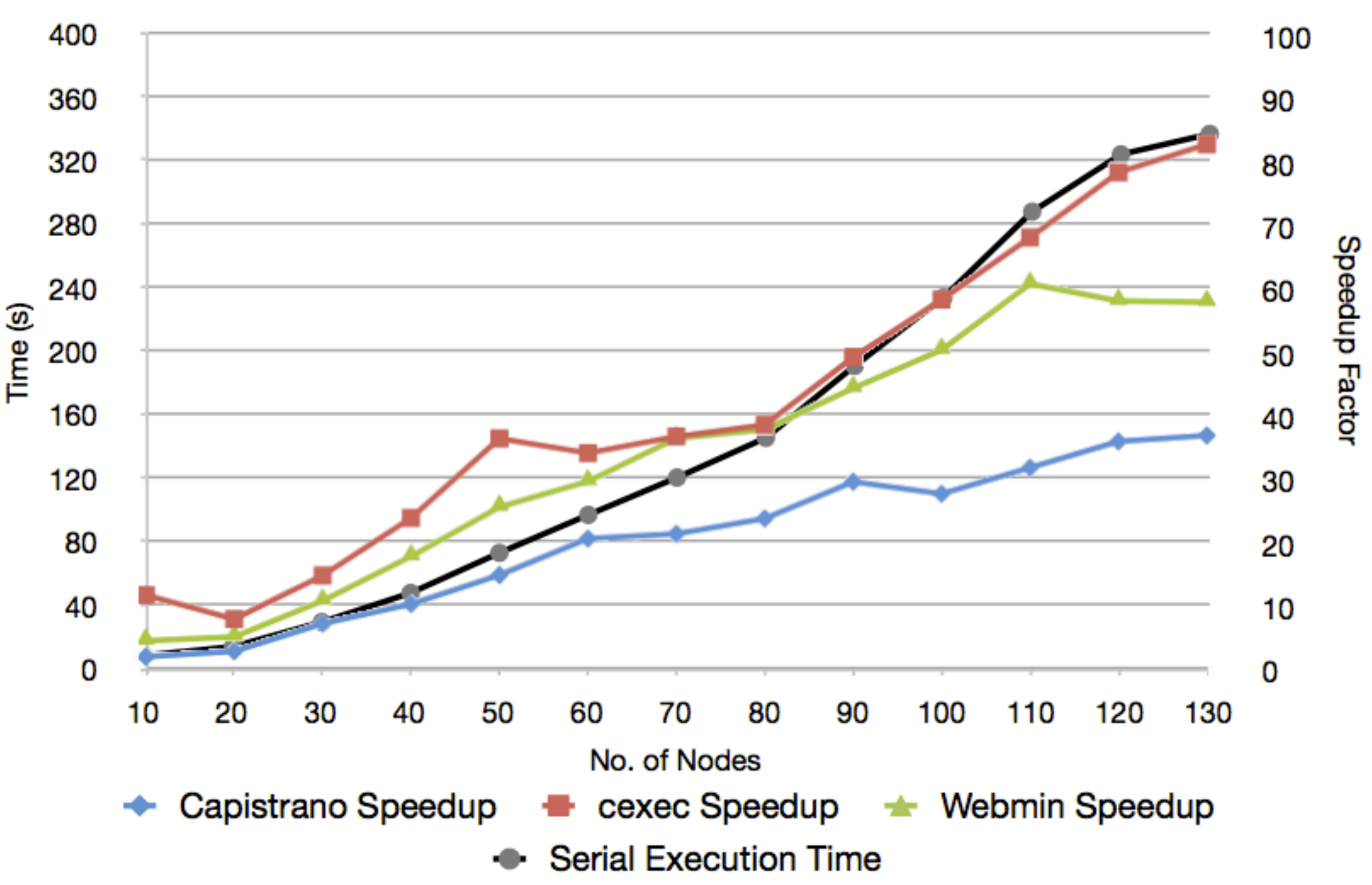}}
\subfloat[]{\includegraphics[width=2.95in]{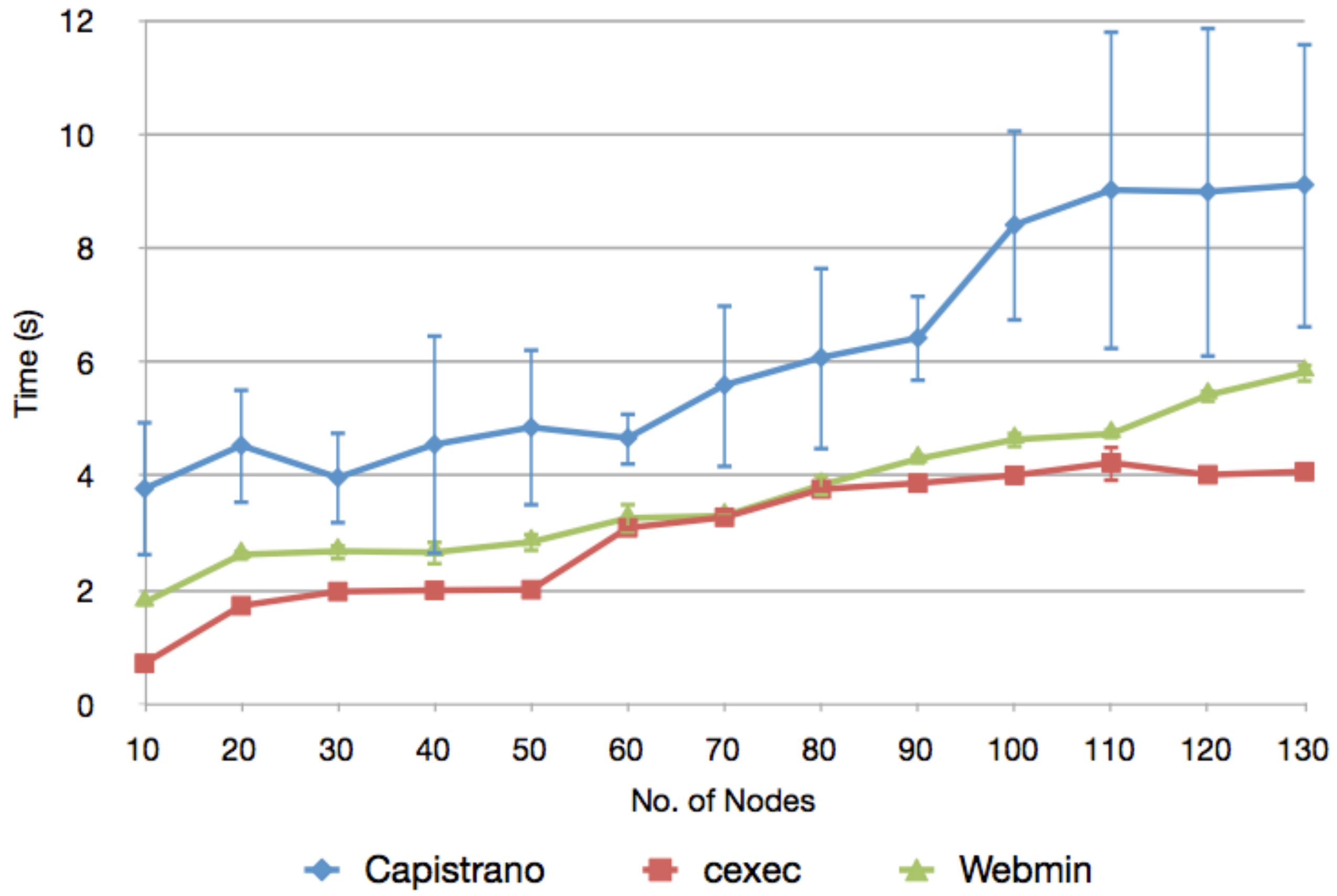}}
\caption{Comparison of Parallel Execution Tools and Speedup}
\end{figure*}

We initially see that the page load times are small using both Ganglia (blue) and the C2MS (red), with the data being displayed in a matter of milliseconds. By using 95\% confidence intervals, we see that any system can potentially produce the same page load times. As the number of servers increase in the cloudlet, the page load times increase slightly however with the exception of the recorded times at 110 servers when using Ganglia. This can be attributed to host performance variation as the C2MS data point at 20 nodes also provides a larger variation than expected. We see that no major time differences of page loads exist between the two tools and this can be attributed to linking to original server data without replication and the low overhead of graph aggregation.
 
\subsubsection{Apache Load}
Secondly we determine if viewing a cloudlet's monitoring output via the C2MS places additional load on the Apache server displaying the data, when compared to regular Ganglia. Upon loading a page 5 times, we recorded the total CPU load placed on the Apache server; this is performed three times (i.e, 15 page loads in total) for each cloudlet size and the average value is taken. We used Apache's Server Status module to obtain the data values recorded. Figure 6(b) shows our results.

As expected, the larger amount of data to display, the greater the percentage of the CPU is utilized. Both tools show similar results until 80 nodes where the C2MS uses slightly more resources after a dip in utilization. The average differences between the two tools becomes more noticeable at 100 nodes and onwards however this difference is negligible. We also display 95\% confidence intervals to show that in most cases, readings from both tool executions will provide similar results. Hence the overhead introduced by the C2MS and our modified version of the \textit{/var/www/ganglia-web/stacked.php} file is in most cases close to, if not zero. Hence administrators familiar with Ganglia should see no additional latencies when using the C2MS on relatively large cloud infrastructures. As such, the performance differences between the C2MS and other monitoring tools will be similar to the differences between Ganglia hence we need not conduct a performance comparison between the C2MS and other monitoring tools. 

 \subsection{Control Performance}
We now investigate if the C2MS can execute administrator-specified commands quickly over a large set of machines. Currently two versions of the control component are available: serial and parallel SSH command execution. The cluster management tools Capistrano, \textit{cexec} and Webmin were tested for providing concurrent command execution functionality. We expect the parallel version to outperform serial execution but by how much? Which management tool gives the greatest performance?

We investigated the time taken for both the serial and parallel versions to execute a simple \textit{uptime} command over an increasing number of servers. Each command is run 5 times per method and the results are averaged as shown in Figure 7. 

Figure 7(a) displays the serial version execution times (black) where single SSH commands are executed one at a time and are in the format \textit{ssh [hostname] [command]}, as well as the speedup achieved for each parallel execution tool. Figure 7(b) shows the actual execution times of each of these tools with 95\% confidence intervals; due to the small variation between runs for the \textit{cexec} and Webmin tools, these are difficult to view.

We see that the \textit{cexec} tool offers the lowest execution time and greatest speedup when executing a command over up to 130 nodes. Webmin follows closely where execution times and speedup equal that of \textit{cexec}'s at some stages. Capistrano being the slowest of the three still offers fast parallel command execution over 130 nodes taking only approximately 9 seconds however with greater variability. In comparison to the parallel version, serial execution obviously produces much larger execution times with the biggest difference of approximately 338 seconds at 130 nodes. Although the greatest speedup achieved is approximately 3.6 times below the ideal, the speedup of 82 at 130 nodes/processors is a vast improvement on the serial version originally employed. By using \textit{cexec} we achieve the greatest performance and least variability of execution times over a varying number of cloud servers; further experimentation would be required to determine the upper limits of \textit{cexec} as well as the other control tools.

\section{Conclusions and Future Work}
In this paper, we have outlined the C2MS; a dynamic server monitoring and control tool designed specifically for organizations who employ server clustering as a foundation of their cloud infrastructure. This allows administrators to create cloudlets that contain servers that may leave and join other cloudlets. Grouping servers into cloudlets is especially useful for administrators who require high availability, scalability and easy management. Furthermore this is especially useful to help understand individual cloudlet demand to judge whether to increase or decrease cloudlet capacity. Current monitoring tools tend to be static meaning any change in the infrastructure setup requires servers to be reconfigured and restarted to allow the monitoring tool to successfully adopt the new setup; an unreasonable task to undertake on medium to large-scale infrastructures.

We have shown how the C2MS allows administrators to define cloudlets as well as easily add and remove servers to and from these groups without the need for server reconfiguration; via our easy to use C2MS web interface  Administrators can then easily monitor individual cloudlets by viewing dynamically aggregated graphs for the many metrics that Ganglia offers. Furthermore, by adding the appropriate module, the C2MS can make cloudlet power usage and CPU temperature monitoring data available. From the experiments we have performed, we have shown that the C2MS offers quick control of servers as well as effective monitoring with little or no overhead compared to the tool it is built upon.

This tool was originally created for use in a software production company and has also successfully been used to monitor servers within an \textit{ad hoc} cloud computing environment where servers are sporadically available and are grouped to perform specific tasks; more information about \textit{ad hoc} cloud computing can be found at \cite{Kirby2010}. The C2MS can not only be used upon clouds but any platform Ganglia can be installed hence system administrators who wish to try a demo of the C2MS as well as download the tool, can find the respective links at \cite{c2ms_gary}.

\section*{Acknowledgements}
This work has been partially funded by ICG Software, the HPC-EUROPA2 project (project number: 228398) with the support of the European Commission Capacities Area - Research Infrastructures Initiative, the EPSRC NeSC Research Platform grant, the Amazon EC2 Research Grant Award and the MEyC-Spain under contract TIN2011-28689-C02-02.

\bibliographystyle{IEEEtran}
\bibliography{Papers_Read,Other_Refs,My_Papers}	

\end{document}